\documentclass[5p, times]{elsarticle}

\usepackage{amsmath}
\usepackage{algorithm}
\usepackage{algorithmic}
\usepackage{graphicx}
\usepackage{lineno,hyperref}
\usepackage{caption}
\usepackage{array} %\调用公式宏包的命令应放在调用定理宏包命令之前，也能控制表格
\usepackage{booktabs} %调整表格线与上下内容的间隔
\usepackage{longtable}%调用跨页表格
\usepackage{natbib} %参考文献管理包
\usepackage{amsthm,amssymb,amsfonts}
\usepackage{wrapfig}
\usepackage{color}

\journal{Journal of \LaTeX\ Templates}

\begin{document}
\captionsetup[figure]{labelfont={bf},name={Fig.},labelsep=period}
\biboptions{sort&compress}%参考文献可以压缩显示例如1-3
\begin{frontmatter}

\title{Content-Based Multi-Source Encrypted Image Retrieval in Clouds with Privacy Preservation}

\author[1staddress]{Meng Shen\corref{mycorrespondingauthor}}
\cortext[mycorrespondingauthor]{Corresponding author}
\ead{shenmeng@bit.edu.cn}

\author[1staddress]{Guohua Cheng}
\ead{chengzi92036@163.com}

\author[1staddress]{Liehuang Zhu}
\ead{liehuangz@bit.edu.cn}

\author[2ndaddress]{Xiaojiang Du}
\ead{dxj@ieee.org}

\author[3rdaddress]{Jiankun Hu}
\ead{J.Hu@adfa.edu.au}

\address[1staddress]{School of Computer Science, Beijing Institute of Technology, Beijing 100081, China}
\address[2ndaddress]{Department of Computer and Information Sciences, Temple University, Philadelphia, PA 19122, USA}
\address[3rdaddress]{School of Engineering and IT, University of New South Wales (UNSW), Canberra, ACT 2610, Australia}

\begin{abstract}

Content-based image retrieval (CBIR) is one of the fundamental image retrieval primitives. Its applications can be found in various areas, such as art collections and medical diagnoses.
With an increasing prevalence of cloud computing paradigm, image owners desire to outsource their images to cloud servers.
In order to deal with the risk of privacy leakage of images, images are typically encrypted before they are outsourced to the cloud, which makes CBIR an extremely challenging task.
Existing studies focus on the scenario with only a single image owner, leaving the problem of CBIR with multiple image sources (i.e., owners) unaddressed.

In this paper, we propose a secure CBIR scheme that supports \underline{M}ultiple \underline{I}mage owners with \underline{P}rivacy \underline{P}rotection (\textsf{MIPP}). We encrypt image features with a secure multi-party computation technique, which allows image owners to encrypt image features with their own keys.
This enables efficient image retrieval over images gathered from multiple sources, while guaranteeing that image privacy of an individual image owner will not be leaked to other image owners.
We also propose a new method for similarity measurement of images that can avoid revealing image similarity information to the cloud. Theoretical analysis and experimental results demonstrate that \textsf{MIPP} achieves retrieval accuracy and efficiency simultaneously, while preserving image privacy.
\end{abstract}


\end{frontmatter}

%\linenumbers

\section{Introduction}
Recent years have witnessed the prosperity of image-sharing services and applications (e.g., Instagram),
which results in an increasing demand for image retrieval.
In early years, text-based image retrieval systems implemented by manual tagging image properties could fulfill the requirement of image retrieval.
With the growing popularity of Internet users, hundreds of millions of images appear on the Internet per second, the traditional text-based image retrieval becomes gradually impractical, because it consumes prohibitive manpower and financial resources for labeling.. Content-based image retrieval (CBIR)~\cite{cbir1,cbir2,cbir3} has been proposed for real world applications which uses the image feature extracted automatically from images, such as colors~\cite{color1,color2}, textures~\cite{texture1,texture2}, and shapes~\cite{shape1,shape2}.

In general, higher resolution images consume more storage.
For instance, a photo taken by a cell phone in present days may be about 2MB and the one taken by a professional camera may reach 10MB or more.
With an increasing prevalence of cloud computing and storage ~\cite{cloud1,cloud2}, migrating services to the cloud has rapidly become a trend for mass data storage and management.
By outsourcing images to the cloud, service providers can make their services easily accessible to geographically distributed users by only requiring them to pay for the computation and storage resources they actually use.

Outsourcing images directly to cloud servers, however, increases the risk of privacy leakage when images contain sensitive information, such as patient's medical information or personal location information.
For instance, a compromised cloud vendor could enable access to the outsourced images by unauthorized users.
In order to protect images against privacy leakage threats, images are usually encrypted before being outsourced to the cloud.
Since encryption operations disrupt the image content, it becomes a challenging task to perform CBIR over encrypted images.
Therefore, it is highly desirable to devise a privacy-preserving CBIR system for cloud-based encrypted image sets.

Many schemes have been proposed in the field of secure CBIR
~\cite{ZhangY,ZhangL,XiaZH,YuanJ,YuanXL,XiaZH1,LiW,FerreiraB,ZhangXP,ChengH,ZhangCY}, which can be roughly classified into two categories.
In the first category, image owners extract features from plain images, and then outsource both the encrypted images and the encrypted image features to the cloud.
In the second category,
image owners outsource only the encrypted images to the cloud that is responsible for extracting features from encrypted images and for conducting retrieval operations.

Existing studies have a common limitation that they consider only a single-source case (i.e., a single image owner).
In real-world applications, however, image retrieval is more likely to get multiple image sources involved.
For instance, consider a cloud-based e-health application,
which takes an encrypted ultrasonic medical image of an undiagnosed patient as input and searches for similar confirmed cases from a collection of encrypted medical images.
Suppose images are collected from multiple hospitals (i.e., sources), each of which is reluctant and unpermitted to share with one another the plain medical images.
The existing schemes can be easily extended to
the multi-source scenario by performing retrieval over encrypted images of different owners \emph{one by one}.
Although simple and straightforward, it introduces multiple rounds of communications between users and individual image owners, and thereby becomes inefficient in terms of retrieval time and communication overhead.

There are several challenges in designing a secure and efficient CBIR scheme with multiple image owners.
First, we should ensure the privacy of images and image features of different image owners.
Second, the authorized query user should communicate his secret image encryption key with image owners for generating a secret query in secure image retrieval schemes. However, when secure image retrieval schemes have multiple image owners, each image owner should use their own secret image encryption key to encrypt images and image features. Then, the authorized query user should communicate his secret image encryption key with each image owner for generating a secret query, which will increase the communication overload in schemes.
It is desirable to address this problem in the secure image retrieval scheme with multiple image owners.
Finally, when the cloud executes image retrieval,
it may obtain similarity relation information of images in the retrieval result. This privacy issue should be also addressed.

In this paper, we propose the \textsf{MIPP}, a novel content-based multi-source image retrieval scheme with privacy protection.
\textsf{MIPP} operates in the same way as existing schemes in the first category, which outsources encrypted images along with their encrypted image features to the cloud.
In order to address the challenges of supporting multiple image owners, we first encrypt images with a key stream and encrypt the corresponding image features by the secure multi-party computation method, and then propose a novel method to measure the image similarity; this can help to avoid revealing the image similarity information in cloud to a certain extent.

The main contributions in this paper are highlighted as follows:
\begin{itemize}
\item We design a \textsf{MIPP}, which, to the best of our knowledge, is the first scheme belonging to the first category that enables content-based multi-source image retrieval with privacy protection.
    In the proposed \textsf{MIPP}, multiple image owners are allowed to encrypt images and image features by their unique secret image encryption keys. This enables an efficient image retrieval over images gathered from multiple sources, while providing guarantees that image privacy of an individual image owner will not be leaked to other image owners.
    Thus, the proposed \textsf{MIPP} can meet the practical requirements in real-world applications.
\item We present a new approach to measure the similarity of images, which can avoid the leakage of image similarity information in retrieval results.
    Extensive experimental results show that the resulting retrieval outcome is comparable to that with the typical Euclidean distance criterion.
\end{itemize}

The rest of this paper is organized as follows.
We summarize the related work in Section \ref{sec:related_work} and introduce the preliminaries in Section \ref{sec:preliminaries}.
In Section \ref{sec:system_model}, we present the system model, thread model, and design goals of our scheme.
We detail the design of \textsf{MIPP} in Section \ref{sec:MIPP} and
present the security analysis in Section \ref{sec:security}.
We evaluate the performance of the proposed scheme in Section \ref{sec:evaluation} and conclude this paper in Section \ref{sec:conclusion}.

\section{Related work}\label{sec:related_work}
In this section, we will present a brief overview of existing research schemes in the field of secure image retrieval.

Homomorphic encryption is a form of encryption that allows computations to be carried out on ciphertext, thus generating an encrypted result which, when decrypted, matches the result of operations performed on the plaintext. Zhang et al.~\cite{ZhangY,ZhangL} leveraged the property of homomorphic encryption in secure image retrieval. The homomorphic encryption results in high computational complexity that makes it consume too much time. Xia et al.~\cite{XiaZH} proposed a secure CBIR scheme with Bag-of-Words model and Earth Mover’s Distance. The retrieval index was constructed by locality-sensitive hashing. In this scheme, the user and image owner have two times of two-way communications, which results in high communication overhead.
Yuan et al.~\cite{YuanJ} proposed a scheme named SEISA with access control and secure k-means outsource, dynamically updating images is supported. Yuan et al.~\cite{YuanXL} proposed a scheme that can explore user relationship while preserving image privacy. The secure index and encrypted image features were constructed by an entity called SF rather the image owner.
This scheme supports dynamic updates of images without affecting the current social structure.
The scheme proposed by Xia et al.~\cite{XiaZH1} was able to deter the illegal distribution of images while preserving the image privacy.
Li et al.~\cite{LiW} proposed a privacy preserving retrieval scheme for outsourced media, which used the one-way hash along with encrypting partly hash values to encrypt image features. This scheme created trade offs among privacy preserving, retrieval quality, and complexity through adjusting the bit counts of encryption in the hash value.

In the second category, Ferreira et al.~\cite{FerreiraB} proposed a scheme IES-CBIR that can extract image features from encrypted images. The texture and color features were encrypted separately; the color feature was encrypted by scrambling pixels in HSV color model and the texture feature is encrypted by shuffling rows and columns in images.
This scheme enabled dynamic updates of images by using Bag-Of-Visual-Words model.
Zhang et al.~\cite{ZhangXP} proposed a histogram-based retrieval scheme for encrypted JPEG images with machine learning.
They encrypted images by permuting DCT and the server to retrieve the histogram at each frequency position from encrypted images. %Moreover, this scheme support multiple image owners.
Cheng et al.~\cite{ChengH} proposed a markov process based retrieval scheme for encrypted JPEG images.
Markov's process models of the AC coefficients and the server could extract features from the transition probability matrices of those AC coefficients of encrypted images.
Zhang et al.~\cite{ZhangCY} proposed an encrypted medical image retrieval algorithm based on DWT-DCT frequency domain. In this algorithm, features were extracted from encrypted images.
%and this scheme has good robustness to against some attacks.

As described in Section 1, these schemes may lead to heavy communication overload and privacy leakage (e.g., image features and image similarity) when simply extended for CBIR with multiple image owners.

\section{Preliminary}\label{sec:preliminaries}
In this section, we will introduce the preliminaries,
including image features used in this paper and the secure multi-party computation.

\subsection{Image feature}
MPEG-7~\cite{ManjunathBS,MejiaLavalle} standard is the multi-media content description interface that contains a
set of descriptors. We extract the Edge Histogram Descriptor
(EHD) from images for our secure CBIR scheme. Edge Histogram Description is a non-homogenous texture descriptor in MPEG-7 which captures spatial distribution of edges and works well in CBIR.

\subsection{Secure multi-party computation}
Assume that there are multiple parties, each of which owns a secret number.
Each party expects to obtain the total number of all parties, without publicizing their own number to others.
Thus, each participant needs to encrypt his number before making it public.
The secure multi-party computation can calculate over encrypted numbers, which meets the above requirement.
Under the background of the secure multi-party computation technology, we can obtain the sum or product of numbers in the same manner as calculating under plain number. Nowadays, secure multi-party computation has been applied in many real-world applications, such as secure voting and secure electronic auction.

Jung et al.~\cite{JungT} proposed a privacy-preserving sum calculation scheme with collusion-tolerable, without the need for secure channels.
This scheme can effectively calculate the sum of encrypted numbers, which can be briefly described as follows.

\textbf{Step 1}: Select two large prime numbers $p$ and $q$ with the same length where $q$ divides $p-1$.

\textbf{Step 2}: Define the $q$-order cyclic multiplicative group G$_{1}$ with a generator being defined in Equation~\eqref{eq:eq1}, where $h$ is a random number in Z$_{p}$.
And define the $q$-order multiplicative group G$_{2}$, where its generator is defined in Equation~\eqref{eq:eq2}.
\begin{equation}
g_1 = {h^{(p-1)/q}\quad mod\quad p\quad s.t.\quad g_1 \not= 1\quad mod\quad p}
\label{eq:eq1}
\end{equation}
\begin{equation}
g_2 = {g_1^p\quad mod\quad p^2}
\label{eq:eq2}
\end{equation}

\textbf{Step 3}: Each participant $P_{i}$ randomly chooses a number $r_{i}$ $\in$ Z$_{q}$ and calculates a public number ${g_{2}}^{r_{i}} mod {p^{2}}$. Then, she exchanges the number ${g_{2}}^{r_{i}}$ $\in$ G$_{2}$ with $P_{i-1}$ and $P_{i+1}$. $P_{i}$ can calculate a secret number $R_{i}$ $\in$ G$_{2}$ as shown in Equation~\eqref{eq:eq3} and the ciphertext $C_{i}$ as shown in Equation~\eqref{eq:eq4} after a round of exchanges.
\begin{equation}
R_i = {(g_2^{r_{i+1}}/g_2^{r_{i-1}})^{r_i}}
\label{eq:eq3}
\end{equation}
\begin{equation}
C_i = {(1+x_ip)R_i\quad mod\quad p^2}
\label{eq:eq4}
\end{equation}

\textbf{Step 4}: Each $P_{i}$ shares her ciphertext $C_{i}$ with other participants and calculates the product of all ciphertexts according to Equation \eqref{eq:eq5} to obtain the value of $C$.
\begin{align}\label{eq:eq5}
C   &={\prod\limits_{i=1}^{n}C_{i}}\quad mod \quad p^2 \nonumber \\
&   ={\prod\limits_{i=1}^{n}(1+x_{i}p)({g_2^{r_{i+1}}}/{g_2^{r_{i-1}}})^{r_i}}\quad mod \quad p^2 \nonumber\\
&   =(1+p\sum_{i=1}^{n}x_i)g_2^{\sum_{i=1}^{n}r_{i+1}r_{i}-r_{i}r_{i-1}}\quad mod \quad p^2\\
&   =(1+p\sum_{i=1}^{n}x_i)\quad mod \quad p^2 \nonumber
\label{eq:eq5}
\end{align}

\textbf{Step 5}: The sum of all numbers can be obtained by calculating Equation~\eqref{eq:eq6}.
\begin{equation}
(C-1)/p = \sum_{i=1}^{n}x_i\quad mod\quad p
\label{eq:eq6}
\end{equation}

\section{Problem Formulation}\label{sec:system_model}
In this section, we will introduce the system model, threat model, and design goals of our scheme.

\subsection{System model}
\begin{figure*}[t]
\centering
\includegraphics[height=8cm]{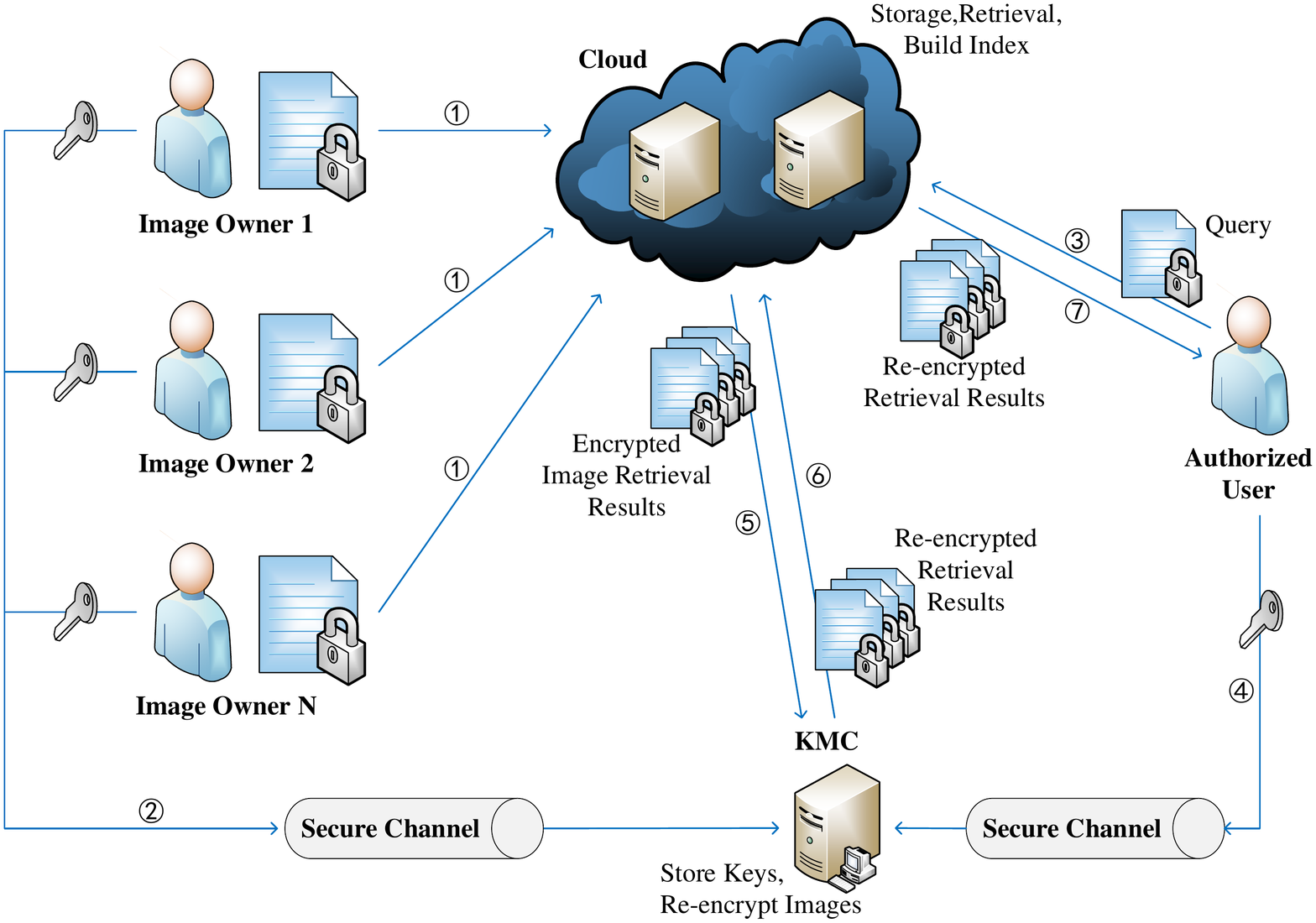}
\caption{System Model of the Secure Multi-Source CBIR Scheme}
\label{fig:sysmodel}
\end{figure*}

There are four types of entities in our secure multi-source CBIR system, including multiple image owners, authorized query users, the cloud, and a key management center (KMC), as illustrated in Fig.~\ref{fig:sysmodel}.
The description of each type of entity is detailed as follows.

\begin{itemize}
\item \textbf{Multiple images owners}: They are providers of image databases denoted by Image Owner $i$ ($i \in [1,N]$) in Fig.~\ref{fig:sysmodel}.
    We assume that each image owner has a secure channel to communicate his secret key with KMC.

\item \textbf{Authorized query users}: They are users authorized by specific image owners and have the authority to send image retrieval requests to the cloud. We assume that authorized query users in the system will not reveal their secret image encryption key or distribute image retrieval results to unauthorized users.

\item \textbf{Cloud}: It takes responsibility for building secure retrieval indexes and executing image retrieval. It stores encrypted images, encrypted image features, information of image owners, and a list of authorized query users.
    We assume that the cloud is honest-but-curious, which means it will execute the image retrieval operation correctly, while it may also analyze images or image features to obtain some sensitive information about images.

\item \textbf{Key management center (KMC)}: It has three main functionalities. First, it takes responsibility for storing secret image encryption keys, information of image owners, authorized query user lists received from image owners, and storing the authorized query user's secret image encryption key temporarily when there comes a query from an authorized query user. Second, it will decrypt encrypted image retrieval results from the cloud, and then encrypt these images with the secret image encryption key of the authorized query user. Finally, it will send the new encrypted image retrieval results to the cloud. In this paper, we assume that the KMC is fully trusted that it will not reveal secret image encryption keys of image owners and authorized query users to others.
\end{itemize}

The workflow of our scheme is described as follows:

\begin{enumerate}[(1)]
\item Multiple image owners extract the EHD feature to represent an image, then they encrypt their images and image features respectively. After that, they outsource encrypted images and encrypted image features along with their identity to the cloud. The authorized user list will also be sent to the cloud for the image retrieval service, shown as \textcircled{1} in Fig.~\ref{fig:sysmodel}. They also need to send secret image encryption keys that is used to encrypt images to the KMC through a secure channel, shown as \textcircled{2} in Fig.~\ref{fig:sysmodel}. Image owners need not store secret keys that is used to encrypt image features. When a new image owner arrives, the above operations should be repeated.

\item Before an authorized user requests a query, he/she will extract EHD features from query images and encrypt query image features, then submit the generated encrypted query to the cloud for image retrieval operation, shown as \textcircled{3} in Fig.~\ref{fig:sysmodel}. At the same time, he should send his secret image encryption key to the KMC through a secure channel, shown as \textcircled{4} in Fig.~\ref{fig:sysmodel}. For each retrieval request, the image encryption key that the authorized query user sends to the KMC should be different from the last time he retrieved. After receiving retrieval results from cloud, the authorized query user decrypts retrieval results with his own image encryption key, which is the same as the key sent to the KMC. Finally, the authorized query user can sort the decrypted retrieval results to get top-h similar images.

\item After receiving encrypted images and image features from image owners, the cloud will build the retrieval index. When an image retrieval request arrives, he should verify the identity of query user. Then, it will execute an image retrieval operation if verified successfully. When it obtains retrieval results, it will send the encrypted retrieval results to the KMC instead of sending the retrieval results to the authorized query user, shown as \textcircled{\small{5}} in Fig.~\ref{fig:sysmodel}. At the same time, information of the authorized query user will also be sent to the KMC along with the corresponding encrypted retrieval results.

\item When the KMC receives encrypted images from cloud, it will decrypt these images with image owners' secret image encryption key first. Then, it will encrypt these images with the secret image encryption key of the related authorized query user. After that, the re-encrypted images will be sent to the cloud, shown as \textcircled{6} in Fig.~\ref{fig:sysmodel}.

\item Cloud will return the re-encrypted image retrieval results that is received from the KMC to the authorized query user, shown as \textcircled{\small{7}} in Fig.~\ref{fig:sysmodel}.

\end{enumerate}

\subsection{Threat model}
In our scheme, we consider the following two kinds of threads:

\begin{enumerate}[(1)]
\item Eavesdroppers.

In the process of image transmission (e.g., sending encrypted images and their features to the cloud, and fetching retrieval results from the KMC, the KMC sending re-encrypted images to the cloud, and the cloud providing retrieval results to the authorized query user), eavesdroppers may eavesdrop image information. This is a weak adversary that can be defended by encryption.

\item Cloud.

In our scheme, we assume the cloud is honest-but-curious. It will correctly execute the image retrieval operation, but it may analyze image content through encrypted images with their features at the same time. Thus, we should guarantee the data privacy in the process of encrypting images and image features. Additionally, the cloud may obtain the similarity relation information of images over the process of secure image retrieval. We should avoid this kind of image privacy leakage in cloud.

\end{enumerate}

\subsection{Design goals}
In this section, we describe design goals of our scheme as follows.
\begin{enumerate}[(1)]
\item Image privacy.

Image privacy is very important in secure image retrieval service. We should unable the cloud and unauthorized users to obtain plain images and plain image features, along with image similarity relation information through encrypted images, encrypted image features, the encrypted retrieval results.

\item Retrieval accuracy.

The retrieval accuracy is an indispensable element in secure image retrieval. In our scheme, we proposed a new approach to manage the similarity of images. Therefore, retrieval accuracy difference between our scheme and the secure scheme retrieval with Euclidean distance should be within reasonable limits.

\item Efficiency.

Efficiency indicates the time consumption in a secure image retrieval scheme. We should ensure high efficiency in our scheme to enhance its practicality in real world application, which means the time of encryption, index construction, and retrieval should be reduced.
\end{enumerate}

\section{The design of \textsf{MIPP}}\label{sec:MIPP}
In this section, notations in our scheme are shown in the table. We first introduce the system overview and then describe the data encryption method.
In order to solve the image similarity leakage problem in the cloud, we propose a new method to measure image similarity.
The secure content-based image retrieval, index construction, and index update method of our scheme are also introduced in this section.

\subsection{Notations in this section}
Notations used in our scheme are described in Table 1.

\begin{table*}[!t]
\footnotesize
\center
\renewcommand{\arraystretch}{1.1}
\tabcolsep 15pt %space between two columns. 用于调整列间距
\begin{tabular*}{13cm}{ll}
\multicolumn{2}{l}{\small{\textbf{Table 1}}}\\
\multicolumn{2}{l}{\small{Notations Used in Our Scheme.}}\\\specialrule{0.05em}{3pt}{3pt}
Notation & Description  \\\hline
  $\emph{n}$ & The size of images and image features \\
  $\emph{u}$ & The size of query image and image features. \\
  $\emph{OID}$ & The identity of each image owner\\
  $\emph{AUL}$ & The authorized user list  \\
  $\emph{SK}$ & The secret image encryption key of each image owner\\
  $\emph{USK}$ & The secret image encryption key of query user \\
  $\emph{UID}$ & The user id \\
  $\emph{AK}$ &  The authentication key to  verify query user's identity\\
  $\emph{W} = \{\emph{w$_{1}$}, \emph{w$_{2}$}, ... , \emph{w$_{n}$}\}$ & The plain image collection of each image owner\\
  $\emph{F} = \{\emph{f$_{1}$}, \emph{f$_{2}$}, ... , \emph{f$_{n}$}\}$ & The plain image feature collection of each image owner\\
  $\emph{FF} = \{\emph{f$_{1}$$^{2}$}, \emph{f$_{2}$$^{2}$}, ... , \emph{f$_{n}$$^{2}$}\}$ & The square image feature collection of each image owner\\
  $\emph{EW} = \{\emph{ew$_{1}$}, \emph{ew$_{2}$}, ... , \emph{ew$_{n}$}\}$ & The encrypted image collection of each image owner\\
  $\emph{EF} = \{\emph{ef$_{1}$}, \emph{ef$_{2}$}, ... , \emph{ef$_{n}$}\}$ & The encrypted image feature collection of each image owner\\
  $\emph{EFF} = \{\emph{ef$_{1}$$^{2}$}, \emph{ef$_{2}$$^{2}$}, ... , \emph{ef$_{n}$$^{2}$}\}$ & The encrypted square image feature collection of each image owner\\
  $\emph{QW} = \{\emph{qw$_{1}$}, \emph{qw$_{2}$}, ... , \emph{qw$_{u}$}\}$ & The query image collection of each query user \\
  $\emph{QF} = \{\emph{qf$_{1}$}, \emph{qf$_{2}$}, ... , \emph{qf$_{u}$}\}$ & The query image feature collection of each query user \\
  $\emph{EQ} = \{\emph{eq$_{1}$}, \emph{eq$_{2}$}, ... , \emph{eq$_{u}$}\}$ & The encrypted query image feature collection of each query user \\
  $\emph{QWW} = \{\emph{qf$_{1}$$^{2}$}, \emph{qf$_{2}$$^{2}$}, ... , \emph{qf$_{u}$$^{2}$}\}$ & The square query image feature collection of each query user \\
  $\emph{EQQ} = \{\emph{eq$_{1}$$^{2}$}, \emph{eq$_{2}$$^{2}$}, ... , \emph{eq$_{u}$$^{2}$}\}$ & The encrypted square query image feature collection of each query user \\
  $\emph{S} = \{\emph{s$_{1}$}, \emph{s$_{2}$}, ... , \emph{s$_{h}$}\}.$ & The top-h retrieval results\\
  $\emph{ER} = \{\emph{er$_{1}$}, \emph{er$_{2}$}, ... , \emph{er$_{h}$}\}$ & The top-h encrypted retrieval results \\
  $\emph{NER} = \{\emph{ner$_{1}$}, \emph{ner$_{2}$}, ... , \emph{ner$_{h}$}\}$ & The re-encrypted retrieval results \\
  \hline
\end{tabular*}
\label{tab:notations}
\end{table*}

\subsection{System overview}
There are four entities in our system, each with its own responsibility, described as follows.
\begin{itemize}
\item \textbf{Image owner} is the provider of image database. Each image owner has his own image collection \emph{W}. He will extract EHD image features from \emph{W} and get the image feature collection \emph{F} and \emph{FF}. For preserving the privacy of images and image features, he will generate a secret image encryption key \emph{SK} and run the \textbf{ImageEnc} algorithm to get encrypted image collection \emph{EW} first. Then, he should generate secret image feature encryption keys and run the \textbf{ImageFeatureEnc} process to get the encrypted image feature collection \emph{EF} and encrypted collection \emph{EFF}. Then \emph{EW}, \emph{EF} and \emph{EFF} will be outsourced to the cloud along with \emph{OID}, \emph{AUL}  and \emph{AK} through the network. Besides, \emph{SK} will also be sent to the KMC for storage through a secure channel.
\item \textbf{The authorized user} is the user who has the demand of image retrieval. He will extract EHD features from query images \emph{QW} to obtain the query image feature collection \emph{QF} and run \textbf{ImageFeatureEnc} process to get the encrypted image feature collection \emph{EQ} firstly. Next, he will make calculations to obtain the collection \emph{QWW} and \emph{EQQ}. After that, he will generate a secret query \emph{Q} = \{\emph{EQ}, \emph{EQQ}, \emph{UID}, \emph{AK}\} and send  \emph{Q} to the cloud for secure image retrieval. Finally, he will generate a secret image encryption key \emph{USK} and send this key to KMC through a secret channel. After receiving retrieval results from cloud, he uses his own secret image encryption key \emph{USK} and runs the \textbf{ImageDec} algorithm to decrypt the result images \emph{ER}, then sort the result images to obtain top-h similar images \emph{S} = \{\emph{s$_{1}$}, \emph{s$_{2}$}, ... , \emph{s$_{h}$}\}.

\item \textbf{The cloud} takes responsibility for storing and retrieving. It will store \emph{EW}, \emph{EF}, \emph{OID}, \emph{AUL} and \emph{AK}. Given a query, it uses \emph{AK} to verify query user's identity in the AUL. If validated successfully, it will run the \textbf{ImageRetrieval} process to retrieve similar images in the image database. Instead of sending top-h encrypted retrieval results \emph{ER} to authorized query user directly, he sends \emph{ER} and \emph{AK} to KMC firstly. After that, he will send \emph{NER} that is received from \emph{KMC} to the authorized query user. Last, for improving retrieval efficiency, it will run \textbf{IndexConstruct} process to construct the retrieval index \emph{I}.

\item \textbf{The KMC} stores \emph{SK}, \emph{AUL} and temporarily stores \emph{USK}.  After receiving \emph{ER} and \emph{AK} from cloud, it will run the \textbf{ImageDec} algorithm to decrypt \emph{ER} to obtain \emph{W}, then it uses the \emph{AK} to obtain the secret image encryotion key of the authorized query user and runs the \textbf{ImageEnc} algorithm to encrypt \emph{W} with this key to obtain \emph{NER}. Finally, it will send \emph{NER} to the cloud. After it finish the above operations, it can discard the \emph{USK} of this query.
\end{itemize}

\subsection{Data encryption}
For preserving image privacy in the cloud, images and image features should be encrypted before outsourcing to the cloud. We will introduce data encryption methods of our scheme in this section, including key generation, image encryption, image decryption, and image feature encryption method.
\subsubsection{Key generation}
Given a secret parameter $k$, run the \emph{KeyGen(1$^{k}$)} algorithm, so image owners and authorized query users can obtain the secret image encryption key \emph{SK} and \emph{USK} respectively, where the length of \emph{SK} and \emph{USK} is at least the same as the total number pixels in images and the numbers in \emph{SK} and \emph{USK} is between 0 and 255 inclusively.

\subsubsection{ImageEnc\&ImageDec - Image encryption and decryption}
Given a secret key \emph{SK} and an image collection \emph{W}, image owners run the \emph{ImageEnc(SK, W)} algorithm to encrypt images, shown as Algorithm~\ref{alg:enc}. The authorized query user also runs this algorithm to encrypt images with his secret key \emph{USK}. In our image encryption scheme, we use a standard key stream to encrypt images which is secure against the Chosen-Plaintext Attacks (CPA). Thus our image encryption scheme can protect the privacy of image content.

After receiving \emph{NER} from cloud, the user will run the \emph{ImageDec(SK, EW)} algorithm to decrypt images, shown as Algorithm~\ref{alg:dec}. The $M$ and $N$ in algorithm 1 and algorithm 2 are the height and width of images. The KMC also uses this algorithm to decrypt images.
\begin{algorithm}
%\floatname{algorithm}{Algorithm}%更改算法前缀名称
%\renewcommand{\algorithmicrequire}{\textbf{Input:}}% 更改输入名称
%\renewcommand{\algorithmicensure}{\textbf{Output:}}% 更改输出名称
\footnotesize
\caption{ImageEnc(SK, W)}
\label{alg:enc}
\begin{algorithmic}[1]
    \WHILE{$j \neq M$}
        \WHILE{$k \neq N$}
            \STATE $EW_{jk} \Leftarrow SK_{j \times N+k} \oplus W_{jk}$;
            \STATE $k \Leftarrow k+1$;
        \ENDWHILE
        \STATE $j \Leftarrow j+1$;
    \ENDWHILE
\end{algorithmic}
\end{algorithm}

\begin{algorithm}
%\floatname{algorithm}{Algorithm}%更改算法前缀名称
%\renewcommand{\algorithmicrequire}{\textbf{Input:}}% 更改输入名称
%\renewcommand{\algorithmicensure}{\textbf{Output:}}% 更改输出名称
\footnotesize
\caption{ImageDec(SK, EW)}
\label{alg:dec}
\begin{algorithmic}[1]
    \WHILE{$j \neq M$}
        \WHILE{$k \neq N$}
            \STATE $W_{jk} \Leftarrow SK_{j \times N+k} \oplus EW_{jk}$;
            \STATE $k \Leftarrow k+1$;
        \ENDWHILE
        \STATE $j \Leftarrow j+1$;
    \ENDWHILE
\end{algorithmic}
\end{algorithm}
\subsubsection{ImageFeatureEnc - Image feature encryption}
In order to preserve the privacy of image features, image features should be encrypted before they are outsourced to the cloud. Image owners and authorized query users both need to encrypt image features; they use the same method to encrypt image features. For an image feature \emph{f$_{i}$} = \{\emph{a$_{1}$}, \emph{a$_{2}$}, ... , \emph{a$_{l}$}\}, they will first calculate the square of image feature \emph{f$_{i}$} to obtain \emph{f$_{i}$$^{2}$} = \{\emph{a$_{1}$$^{2}$}, \emph{a$_{2}$$^{2}$}, ... , \emph{a$_{l}$$^{2}$}\} where \emph{l} is the dimension of image feature \emph{f$_{i}$}. We use the secure multi-party computation method introduced in Section \emph{3.2} to encrypt image features. The \emph{ImageFeatureEnc} process will be described in detail as follows.

First, similar to secure multi-party computation, they choose \emph{q} as a large primer number, whose length is the same as \emph{p}, satisfies that \emph{q} divided by \emph{p}-1. Then, they select a random number \emph{h} $\in$ Z$_{p}$ and generate the \emph{g$_{1}$} and \emph{g$_{2}$} as Equation~\eqref{eq:eq1} and Equation~\eqref{eq:eq2}.

Second, they randomly choose a number \emph{r$_{j}$} $\in$ Z$_{q}$ and calculate a number \emph{R$_{j}$} as Equation ~\eqref{eq:eq3} for each dimension \emph{a$_{j}$} in image feature \emph{f$_{i}$}.

Finally, they can get the cipertext \emph{ea$_{j}$} of each dimension \emph{a$_{j}$} in \emph{f$_{i}$} by calculating \emph{ea$_{j}$} = (1+\emph{a$_{j}$p})\emph{R$_{j}$} mod \emph{p$^{2}$}. Then, the cipertext of \emph{f$_{i}$$^{2}$} can be obtained by the same way. The encrypted feature and encrypted square feature are shown as follows.

\emph{ef$_{i}$} = \{\emph{ea$_{1}$}, \emph{ea$_{2}$}, ... , \emph{ea$_{l}$}\}
= \{(1+\emph{a$_{1}$p})\emph{R$_{1}$} mod \emph{p$^{2}$}, (1+\emph{a$_{2}$p})\emph{R$_{2}$} mod \emph{p$^{2}$}, ... , (1+\emph{a$_{l}$p})\emph{R$_{l}$} mod \emph{p$^{2}$}\}.

\emph{ef$_{i}$$^{2}$} = \{\emph{ea$_{1}$$^{2}$}, \emph{ea$_{2}$$^{2}$}, ... , \emph{ea$_{l}$$^{2}$}\}
= \{(1+\emph{a$_{1}$$^{2}$p})\emph{R$_{1}$} mod \emph{p$^{2}$}, (1+\emph{a$_{2}$$^{2}$p})\emph{R$_{2}$} mod \emph{p$^{2}$}, ... , (1+\emph{a$_{l}$$^{2}$p})\emph{R$_{l}$} mod \emph{p$^{2}$}\}.

After encrypting all image features, they can obtain the encrypted feature collection \emph{EF} and \emph{EFF}.

It should be brought to attention that the parameter \emph{p} is public to all image owners, authorized query users, and cloud. Parameters \emph{q}, \emph{h}, and \emph{r$_{i}$} in the image feature encryption process can be selected differently among image features. Image owners and authorized users can select by themselves without communicating with others. Additionally, they do not need to store it, which means these parameters can be discarded after using.

\subsection{New approach to manage image similarity}
In the field of image retrieval, the similarity of images is always measured by calculating the distance of image features. If two images are similar, then the distance of them will be very small. Euclidean distance is a type of distance that is typically used to measure the similarity of images, shown as Equation~\eqref{eq:eucdis}. It can calculate the similarity of images accurately. While this will also lead to the problem that when using it to measure the similarity of images in the cloud, the cloud can obtain the image similarity relation information. In order to solve this problem, we propose a new approach to manage image similarity.

Given two image features \emph{X} = \{\emph{x$_{1}$}, \emph{x$_{2}$}, ... , \emph{x$_{l}$}\} and \emph{Y} = \{\emph{y$_{1}$}, \emph{y$_{2}$}, ... , \emph{y$_{l}$}\}, the distance \emph{NewDis} between \emph{X} and \emph{Y} can be calculated as Equation~\eqref{eq:newdis}. Compared with Euclidean distance, we use ${\sum\nolimits_{i=1}^ux_i}/{u}$ and ${\sum\nolimits_{i=1}^uy_i}/{u}$ to replace x$_i$ and y$_i$ respectively in the third part of Equation~\eqref{eq:eucdis}.
\begin{align}
EucDis  &=\sqrt{(x_1-y_1)^2+(x_2-y_2)^2+ ... + (x_u-y_u)^2}\nonumber\\
&   =\sqrt{\sum_{i=1}^ux_i^2 + \sum_{i=1}^uy_i^2 - \sum_{i=1}^u2{x_i}{y_i}}
\label{eq:eucdis}
\end{align}

\begin{align}\label{eq:newdis}
NewDis  &=\sqrt{\sum_{i=1}^ux_i^2 + \sum_{i=1}^uy_i^2 - \sum_{i=1}^u{2\frac{\sum\nolimits_{i=1}^ux_i}{u}\frac{\sum\nolimits_{i=1}^uy_i}{u}}} \nonumber\\
&   =\sqrt{\sum_{i=1}^ux_i^2 + \sum_{i=1}^uy_i^2 - 2u\frac{\sum\nolimits_{i=1}^ux_i}{u}\frac{\sum\nolimits_{i=1}^uy_i}{u}}\\
&   =\sqrt{\sum_{i=1}^ux_i^2 + \sum_{i=1}^uy_i^2 - 2\frac{{\sum\nolimits_{i=1}^ux_i}{\sum\nolimits_{i=1}^uy_i}}{u}}\nonumber
\end{align}

The experimental results show that the accuracy and recall rate of the proposed approach are comparable to the European distance. In addition, the proposed approach can support multi-source encrypted image retrieval. Therefore we used the proposed approach to calculate the similarity between the images in our scheme.

\subsection{Secure content-based image retrieval}
Before requesting a query, authorized query user should generate a secret query \emph{Q} = \{\emph{EQ}, \emph{EQQ}, \emph{UID}, \emph{AK}\}, and then send \emph{Q} to the cloud for image retrieving. After receiving \emph{Q} from authorized query user, the cloud first verifies whether this user is authorized and which owner authorized this user. If validated successfully, the cloud will retrieve in authorized images in the image database through the retrieval index.

For image collection \emph{W} and query image collection \emph{QW}, the similarity of image \emph{w$_{i}$} and image \emph{qw$_{j}$} can be measured by calculating the distance between \emph{f$_{i}$} and \emph{qf$_{j}$}.

However, image collection and image feature collection in the cloud are all encrypted. The similarities between image \emph{w$_{i}$} and image \emph{qw$_{j}$} can be measured by calculating the distance between \emph{ef$_{i}$} and \emph{eq$_{j}$}.

As image features are all encrypted by the secure multi-party computation method, it is the same for the encrypted image feature \emph{ef$_{i}$} = \{\emph{ea$_{1}$}, \emph{ea$_{2}$}, ... , \emph{ea$_{l}$}\} and encrypted query image feature \emph{eq$_{j}$} = \{\emph{eqa$_{1}$}, \emph{eqa$_{2}$}, ... , \emph{eqa$_{l}$}\} where \emph{l} is the dimension of image feature. The distance between them can be calculated according to Equations~\eqref{eq:eq5}, ~\eqref{eq:eq6} and ~\eqref{eq:sim}. We will describe them as follows.

First, the cloud stores the encrypted image feature \emph{ef$_{i}$} = \{\emph{ea$_{1}$}, \emph{ea$_{2}$}, ... , \emph{ea$_{l}$}\} and \emph{ef$_{i}$$^{2}$} = \{\emph{ea$_{1}$$^{2}$}, \emph{ea$_{2}$$^{2}$}, ... , \emph{ea$_{l}$$^{2}$}\}. He will receive the encrypted query image feature \emph{eq$_{j}$} = \{\emph{eqa$_{1}$}, \emph{eqa$_{2}$}, ... , \emph{eqa$_{l}$}\} and \emph{eq$_{j}$$^{2}$} = \{\emph{eqa$_{1}$$^{2}$}, \emph{eqa$_{2}$$^{2}$}, ... , \emph{eqa$_{l}$$^{2}$}\} from an authorized query user. Then, he can get \emph{CEA}, \emph{CEA$^{2}$}, \emph{CEQA}, \emph{CEQA$^{2}$} according to Equation~\eqref{eq:eq5}.

Second, we can obtain the $\sum_{i=1}^{l}ea_i$, $\sum_{i=1}^{l}ea_i^2$, $\sum_{i=1}^{l}eqa_i$ and $\sum_{i=1}^{l}eqa_i^2$, shown as follows.
\begin{equation}
(CEA-1)/p =\sum\nolimits_{i=1}^{l}ea_{i}\quad mod\quad p\nonumber
\end{equation}
\begin{equation}
(CEA^2-1)/p =\sum\nolimits_{i=1}^{l}ea_{i}^2\quad mod\quad p\nonumber
\end{equation}
\begin{equation}
(CEQA-1)/p =\sum\nolimits_{i=1}^{l}eqa_{i}\quad mod\quad p\nonumber
\end{equation}
\begin{equation}
(CEQA^2-1)/p =\sum\nolimits_{i=1}^{l}eqa_{i}^2\quad mod\quad p\nonumber
\end{equation}

Finally, the distance \emph{Sim} between \emph{ef$_{i}$} and \emph{eq$_{j}$} can be obtained by Equation~\ref{eq:sim}. The similarity of images \emph{w$_{i}$} and image \emph{qw$_{j}$} can be measured by this distance value, and the small distance value indicates that they are similar.

\begin{equation}
Sim=\sqrt{\sum_{i=1}^lea_i^2 + \sum_{i=1}^leqa_i^2 - 2\frac{{\sum\nolimits_{i=1}^lea_i}{\sum\nolimits_{i=1}^leqa_i}}{l}}
\label{eq:sim}
\end{equation}

However, there exists a problem during the distance calculation process. Ciphertexts in encrypted image features are very large and the computation complexity of calculating the sum is high, which will consume too much time. Therefore, constructing a retrieval index is very necessary for high retrieval efficiency.

\subsection{Index construction}
We will calculate the sum of encrypted elements $\sum_{i=1}^{l}ea_i$ and $\sum_{i=1}^{l}ea_i^2$ in encrypted features by using the secure multi-party computation during the distance computation process of image features. However, cipertext $ea_i$ and $ea_i^2$ in the encrypted feature are very large, which will results in the sum calculation operation having high computation complexity and consuming too much time. Therefore, we should build a retrieval index to improve retrieval efficiency.
Because the time mainly consumes in computing the sum of cipertext in encrypted features, given an encrypted image feature \emph{ef$_{i}$} = \{\emph{ea$_{1}$}, \emph{ea$_{2}$}, ... , \emph{ea$_{l}$}\}, we can calculate $\sum_{i=1}^{l}ea_i$, $\sum_{i=1}^{l}ea_i^2$ in advance and then store them in the retrieval index table, shown as Table 2.

Given an encrypted query image feature \emph{eq$_{j}$} = \{\emph{eqa$_{1}$}, \emph{eqa$_{2}$}, $\ldots$, \emph{eqa$_{l}$}\}, the cloud only needs to calculate $\sum_{i=1}^{l}eqa_i$, $\sum_{i=1}^{l}eqa_i^2$ and then get $\sum_{i=1}^{l}ea_i$, $\sum_{i=1}^{l}ea_i^2$ from index table when executing encrypted image retrieval. The sum of ciphertexts in encrypted feature is computed in advance which will save much time in the retrieval process and the retrieval efficiency of our scheme is improved.

\begin{table*}[!t]
\footnotesize
\center
\renewcommand{\arraystretch}{1.1}
%\caption{Retrieval index}\label{tab:retrieval_index} \centering
\tabcolsep 15pt %space between two columns. 用于调整列间距
\begin{tabular*}{12cm}{llll}
\multicolumn{4}{l}{\small{\textbf{Table 2}}}\\
\multicolumn{4}{l}{\small{Retrieval index.}}\\
\specialrule{0.05em}{3pt}{3pt}
Image Owner & Image ID & $\sum_{i=1}^{l}ea_{i}$ & $\sum_{i=1}^{l}ea_{i}^{2}$ \\\hline
  $OID_1$ & $OID_1\_Image_1$ & $OID_1\_Image_1\_Sum_1$ & $OID_1\_Image_1\_Sum_2$\\
  $OID_1$ & $OID_1\_Image_2$ & $OID_1\_Image_2\_Sum_1$ & $OID_1\_Image_2\_Sum_2$\\
  $OID_2$ & $OID_2\_Image_1$ & $OID_2\_Image_1\_Sum_1$ & $OID_2\_Image_1\_Sum_2$\\
  ... & ... & ... & ...\\
  $OID_n$ & $OID_n\_Image_1$ & $OID_n\_Image_1\_Sum_1$ & $OID_n\_Image_1\_Sum_2$\\
  \hline
\end{tabular*}
\label{tab:index}
\end{table*}

\subsection{Images and Index update in cloud}
Sometimes image owners may add images to the cloud or delete images from the cloud. When, images in the cloud are changed, image features will also be changed. Images in the retrieval index should be in accord with images in the cloud, therefore the retrieval index should be modified when images in the cloud are changed. Our scheme supports the dynamic updating of images and index in the cloud including update, delete, and add operation.

\begin{enumerate}[(1)]

\item Add operation.

When an image owner requests the cloud to add some images for him, he should send encrypted images and encrypted image features to the cloud. Then, cloud will store these new images and image features in the image database. After that, the cloud will calculate the related data $\sum_{i=1}^{l}ea_{i}$ and $\sum_{i=1}^{l}ea_{i}^{2}$ of encrypted image features and add these two data along with $OID$, image id into the index table.

\item Delete operation.

When an image owner wants to delete images in the cloud, he should send the image IDs to the cloud. If there are image IDs that belong to this image owner, the cloud will delete these images from the index table. The cloud will also delete encrypted images and the corresponding encrypted image features.

\item Update operation.

For preserving the image privacy, an image owner may re-encrypt his images, image features, and then update these images in the cloud. When an image owner requests to update images, the cloud should delete the stale encrypted images and image features using the image IDs. Then, the cloud adds these re-encrypted images and their image features into the image database. Since updating encrypted image features will not change the $\sum_{i=1}^{l}ea_{i}$ and $\sum_{i=1}^{l}ea_{i}^{2}$, there is no need to update the index table.
\end{enumerate}

\section{Security analysis}\label{sec:security}
In this section, we will analyze the security of our scheme, including data privacy and image similarity leakage in the cloud.
\subsection{Data privacy}
In our scheme, data privacy contains the privacy of image content, image features and image similarity information in cloud. We will analyze these three kinds of privacy in the following subsections.

\subsubsection{Image Content Privacy}
As described in Section 5.2.2, the image owner generates a secret key \emph{SK} to encrypt images. The image owner does not want unauthorized user(e.g., cloud, adversary or others) to obtain his image content; he will not reveal his secret key to unauthorized user. In our scheme, the image owner needs to store his secret key in the KMC. We assume the KMC is fully trusted and will not reveal secret keys to unauthorized user. At the same time, we assume the authorized query user will not reveal his secret image encryption key to unauthorized user and will not send image retrieval results to unauthorized user. For the privacy of images outsourced in the cloud, our scheme supports dynamically updating images in the cloud, which means that image owners can re-encrypt images and then outsource these re-encrypted images to the cloud to replace the stale encrypted images. This operation further enhances the privacy protection of images in the cloud. Once an unauthorized user obtains the key of an image owner, he can only crack images of this image owner for a certain period of time, and if the image owner updates his encrypted images in the cloud, this unauthorized user will be unable to crack these re-encrypted images. For the privacy of the image retrieval results, the authorized query user should send a secret image encryption key that is different from his previous query to the KMC for each of his query. Once an unauthorized user obtain the secret image encryption of an authorized query user, he can only crack retrieval results of this user for this query time which enhance the image privacy protection. Since image owners, authorized query users and the KMC will not reveal the secret image encryption key to others, unauthorized users are unabled to obtain secret image encryption keys and they are also unable to obtain plain image content through image owners, authorized query users or the KMC.

In the existing researches, the cloud returns encrypted retrieval results to authorized query users directly. If we also adopt this strategy, once the unauthorized user obtains one image owner's secret image encryption key, then he can brute force encrypted image retrieval results to obtain the plain image contents. This operation is valid through every query operation of different authorized query users. Even though we have illustrated that the unauthorized user is unable to obtain secret image encryption keys, this strategy may also contain some insecure aspects. Therefore, we designed a different strategy. The cloud first sends \emph{ER} to the KMC after it finishing the retrieval operation. The KMC will decrypt all images received from the cloud and encrypt these images with the secret image encryption key of the related authorized query user, and then send these re-encrypted images \emph{NER} to the cloud. Finally, the cloud sends \emph{NER} that are returned by the KMC to the authorized query user. This new strategy can solve the above problem. Images returned by the cloud are encrypted by the secret image encryption key of the authorized query user, even an unauthorized user obtains secret image encryption keys of image owners, he is unable to crack these images correctly. Even though the unauthorized user obtains the secret image encryption key of an authorized query user, he can only access to crack images that are returned to this authorized query user in this query. Furthermore, the authorized query user will send a different image encryption key to the KMC for each query. Even the unauthorized user can crack retrieval results of this query, he is unable to crack retrieval results in the next query using the same key. In our scheme, the unauthorized users are unabled to obtain secret keys of image owners and authorized query users, therefore our scheme can guarantee that image content privacy is unable to be captured by unauthorized user.

\subsubsection{Image feature privacy}

We use the sum protocol in secure multi-party computation model proposed in~\cite{JungT} to encrypt image features. Jung et al.~\cite{JungT} proposed three security models in his paper, shown as Definition 1, 2 and 3.

\textbf{Definition 1} (CDH problem in $G$)\quad The Computational Diffie-Hellman problem in a multiplicative group $G$ with generator $g$ is defined as follows: given only $g$, $g^a$, $g^b$ $\in$ $G$ where $a$, $b$ $\in$ $Z$, compute $g^{ab}$ without knowing $a$ or $b$.

\textbf{Definition 2} (DDH problem in $G$)\quad The Computational Diffie-Hellman problem in a multiplicative group $G$ with generator $g$ is defined as follows: given only $g$, $g^a$, $g^b$, $g^c$ $\in$ $G$ where $a$, $b$, $c$ $\in$ $Z$, decide if $g^{ab}$ = $g^c$.

\textbf{Definition 3} (CDH-Security in $G$)\quad We say our privacy preserving (sum or product) calculation is CDH-secure in $G$ if any Probabilistic Polynomial Time Adversary (PPTA) who cannot solve the CDH problem with non-negligible chance has negligible chance to infer any honest participant’s private value in $G$, i.e., any PPTA’s probability to solve the CDH problem $\epsilon$ satisfies $\epsilon$ $<$ $|$1/p($\kappa$)$|$ for any polynomial p($\cdot$) where $\kappa$ is the order of the group $G$ defined in the CDH problem.

In Jung's paper, each participant $P_i$ will receive the $g_2^{r_{i-1}}$ and $g_2^{r_{i+1}}$ sent from participant $P_{i-1}$ and $P_{i+1}$; therefore, the unauthorized user may obtain $g_2^{r_{i-1}}$, $g_2^{r_{i+1}}$ and $g_2^{r_{i}}$. They has proved that their sum protocol is CDH-secure in $G_2$ in this condition.

\textbf{Theorem 6.1} \emph{Our scheme can protect image feature privacy from being captured by cloud and unauthorized users.}

\emph{Proof:} In our scheme, image owner generates $g_2^{r_{i}}$ for each dimension. The cloud and unauthorized users are enabled
to obtain $g_2^{r_{i-1}}$, $g_2^{r_{i+1}}$ and $g_2^{r_{i}}$, thus our image feature privacy is also CDH-secure in $G_2$.

For an image feature \emph{f$_{i}$} = \{\emph{a$_{1}$}, \emph{a$_{2}$}, ... , \emph{a$_{l}$}\}, the image owner will calculate the \emph{f$_{i}$$^{2}$} = \{\emph{a$_{1}$$^{2}$}, \emph{a$_{2}$$^{2}$}, ... , \emph{a$_{l}$$^{2}$}\} and encrypt them to obtain \emph{ef$_{i}$} and \emph{ef$_{i}$$^{2}$}. The cipertext \emph{a$_{i}$} and \emph{a$_{i}$$^{2}$} are shown as follows:
\begin{equation}
\emph{ea$_{i}$} = (1+\emph{a$_{i}$p})\emph{$R_i$}\quad mod\quad \emph{p$^{2}$}\nonumber = (1+\emph{a$_{i}$p})\emph{$g_2^{(r_{i+1}/r_{i-1})r_i}$}\quad mod\quad \emph{p$^{2}$}\nonumber
\end{equation}
\begin{equation}
\emph{ea$_{i}$$^{2}$} = (1+\emph{a$_{i}$$^{2}$p})\emph{$R_i$}\quad mod\quad \emph{p$^{2}$} = (1+\emph{a$_{i}$$^{2}$p})\emph{$g_2^{(r_{i+1}/r_{i-1})r_i}$}\quad mod\quad \emph{p$^{2}$} \nonumber
\end{equation}

If the unauthorized user wants to obtain a$_{i}$, then he has to solve the secret parameter $R_i$. Because the calculation process of $R_i$ is unknown to him, he is unable to solve $R_i$, and he cannot obtain the plaintext a$_{i}$.

At the same time, the parameter collection $PC$ = \{$g_2$, \{$r_i$\}\} can be chosen among different image features. This means if we want to encrypt image feature $f_1$, $f_1^2$, $f_2$, $f_2^2$, we can choose four different parameter collections $PC$ to encrypt them. Even if unauthorized users obtain one parameter collection $PC$, he is only able to decrypt one image feature. However, unauthorized users are unable to get the parameter collection $PC$. Moreover, we can discard the parameter collection $PC$ directly, as there is no need to store it.

Furthermore, our scheme supports the update of image features in cloud, which enables image owners update their encrypted image features at any time. Once an unauthorized user obtains one parameter collection $PC$, he can only crack one or more encrypted image features that are encrypted by this parameter collection $PC$ and if image owners update their encrypted image features in the cloud then the parameter collection $PC$ that this unauthorized user obtains will be invalidation. This unauthorized user need to obtain the new parameter collection $PC$ to crack encrypted image features. However, image owners will not store the parameter collection $PC$ and they will not reveal the $PC$ to unauthorized users, so unauthorized users will not obtain the $PC$ and they will be unable to crack the plaintext of image features.

According to the above description, unauthorized users are unable to obtain the plaintext of image features. Thus, our scheme can protect the image feature privacy from being captured by unauthorized users.

\subsection{Image similarity leakage in cloud}
In our scheme, we assume that the cloud is `'honest but curious", which means that he will execute the image retrieval operation accurately and at the same time he will analyst the relation or other information of images.
In current research, there always exists the image similarity leakage problem in the cloud during the cloud executing image retrieval operation. Images in the retrieval result are arranged by the similarity to query image, which will reveal the similarity information of images to the cloud.

\begin{figure}[!t]
\centering
\includegraphics[height=180px]{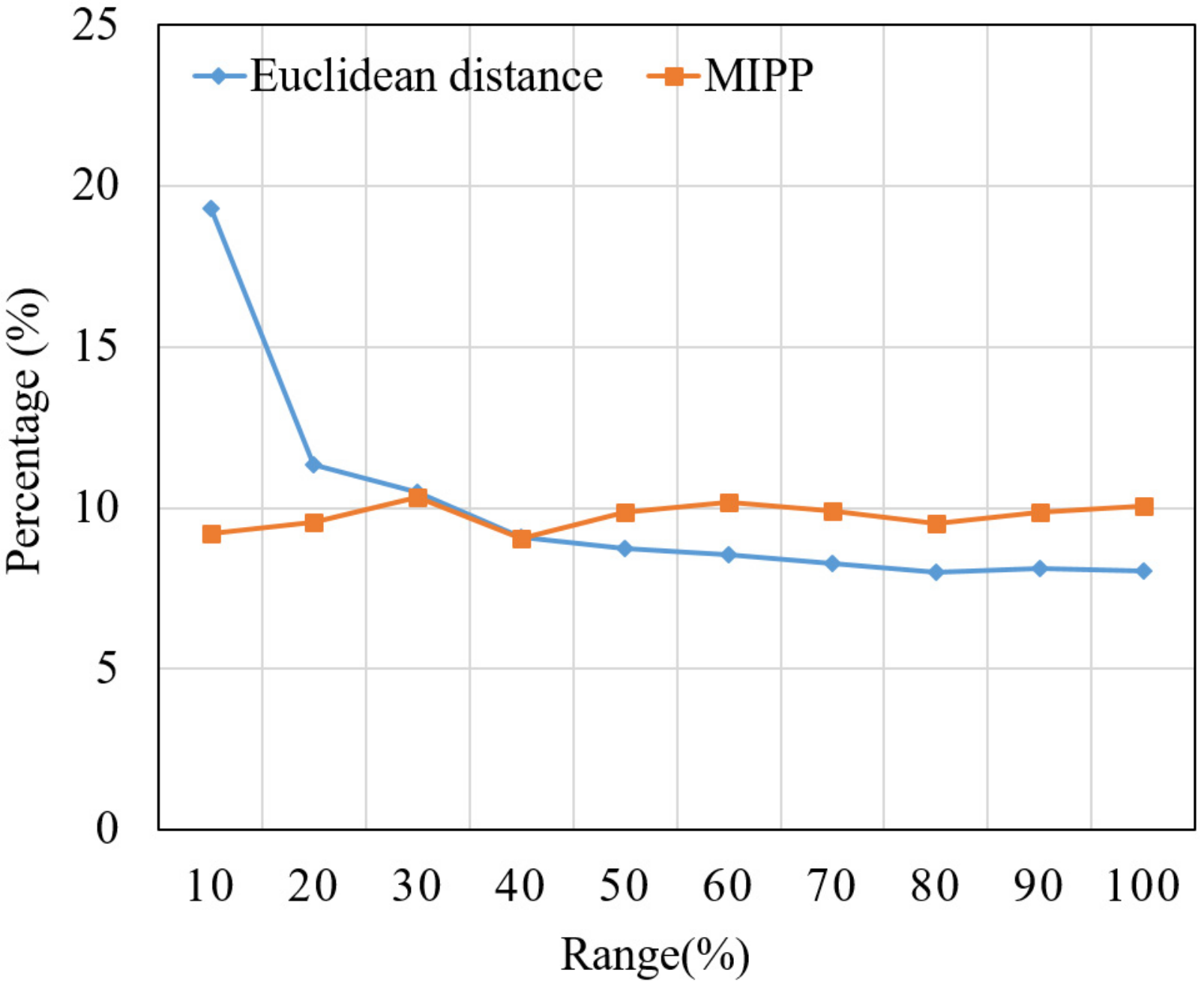}
\caption{The percentage that images will appear in each range of all returned images.}
\label{fig:probability}
\end{figure}

A new distance to manage the similarity of images is proposed in our scheme, which can solve the above problem. Fig.~\ref{fig:probability} shows the distribution of images that is similar to query images in retrieval results when the cloud returns top 100 images. The abscissa is the percentage that similar images distribute in the retrieval results, and the ordinate is the probability that similar images distribute in the designated range. We can see that when retrieving with the Euclidean distance, the percentage of truly similar images appear in the top 10$\%$ of all retrieval result images, which is very high, and the distribution percentage is decreasing from beginning to end in retrieval results. The truly similar image distribution of our scheme is uniform and similar images are not likely to distribute at the beginning percentage of the retrieval result, which will mislead the cloud to analyst the similarity of the images. Because similar images are uniformly distributed in the retrieval results, the cloud may get the wrong similarity relation. Therefore, our scheme can prevent the image similarity leakage to the cloud.

\section{Performance evaluation}\label{sec:evaluation}
In this section, we will introduce the performance evaluation of our scheme, including experimental setting, evaluation of retrieval accuracy, evaluation of time consumptions, and evaluation of storage consumption.

\subsection{Experimental setting}
The corel images~\cite{corel1,corel2} data set is usually used to verify the experiment scheme in the research field of image retrieval. It contains 100 categories each of which has 100 images. They are selected as test images in our experiments.
We choose 10 categories and generate 5 queries for each category, so there are 50 queries in total to evaluate the retrieval accuracy. The proposed scheme is implemented by C++ on Intel Core(TM) Processor 2.7 GHZ.

\subsection{Evaluation of retrieval accuracy}
\begin{figure}[!t]
\centering
\includegraphics[height=180px]{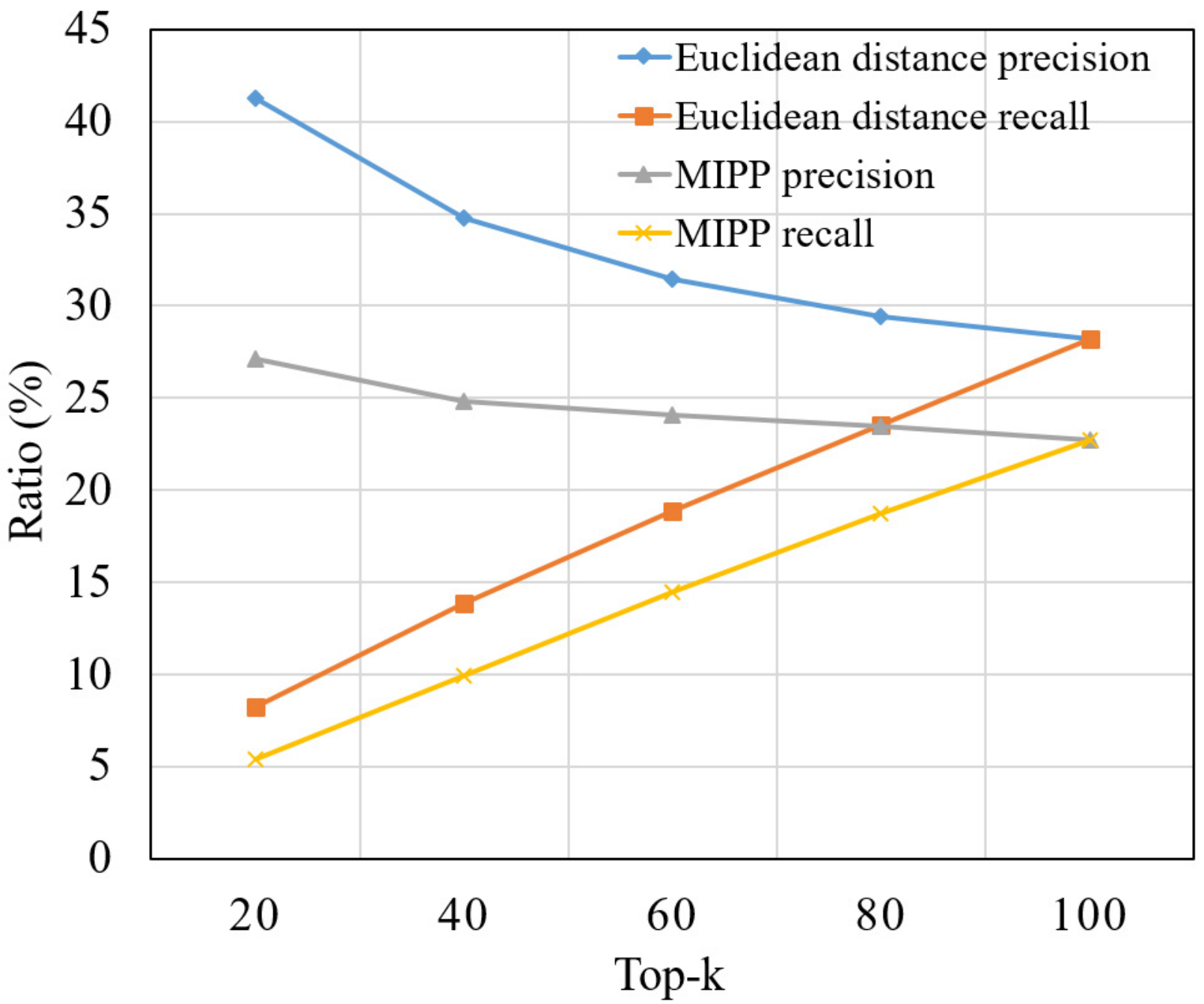}
\caption{Retrieval accuracy of our scheme and Euclidean distance scheme.}
\label{fig:accuracy}
\end{figure}

\begin{figure}[!t]
\centering
\includegraphics[height=180px]{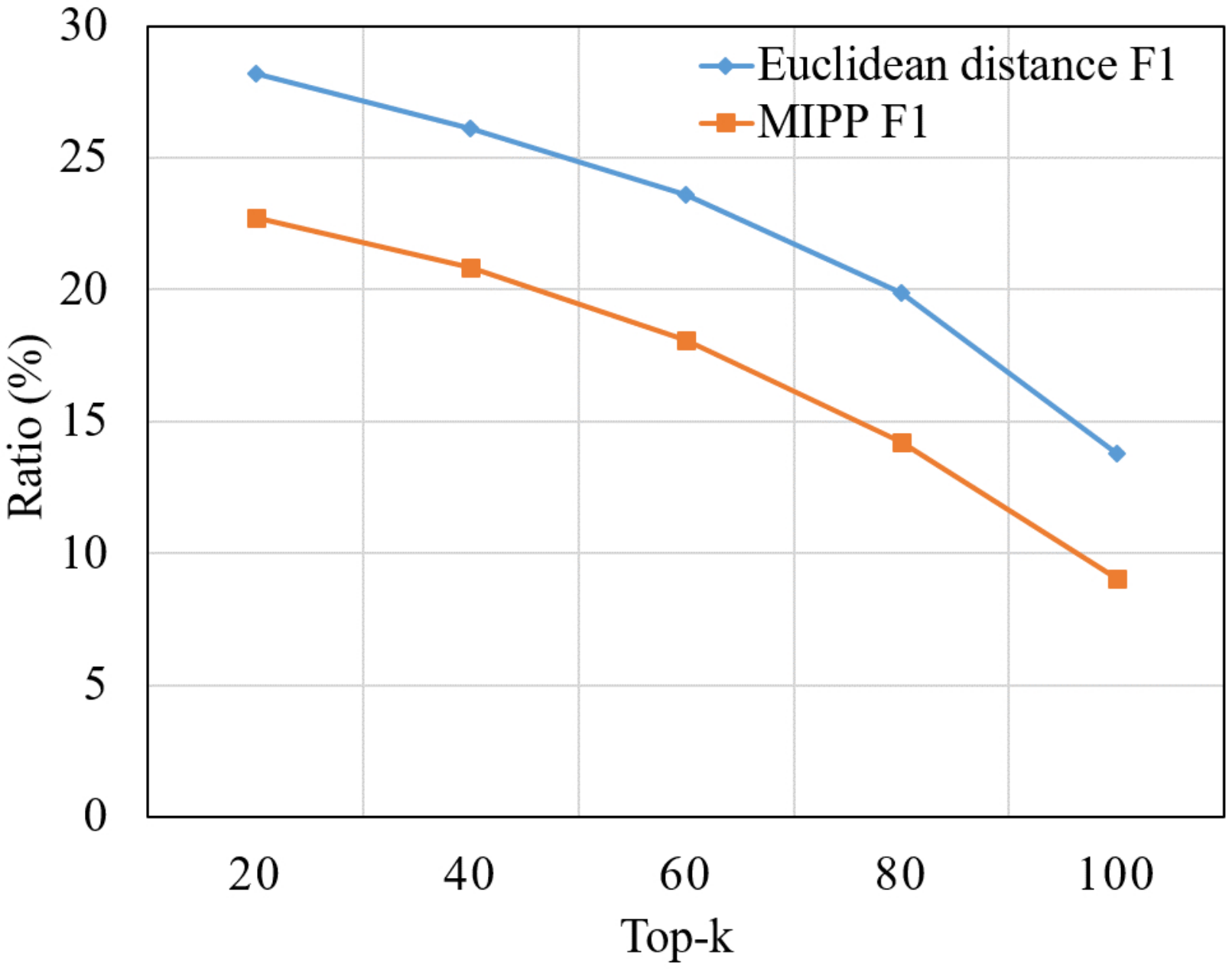}
\caption{F1-Measure of our scheme and Euclidean distance scheme.}
\label{fig:f1}
\end{figure}

In the field of information retrieval, precision, recall ratio, and F1-Measure are typical metrics to evaluate the retrieval results as formulated in Equations~\eqref{eq:precision}-~\eqref{eq:F}, where TP represents the true positives, FP represents the false positives, and FN represents the false negatives.
The precision and recall always represent the contrary variation tendency as shown in Fig.~\ref{fig:accuracy}.
We can use the F1-Measure to make a comprehensive evaluation.
Fig.~\ref{fig:accuracy} shows that the retrieval accuracy of our scheme is about 10\% lower than the scheme retrieval by Euclidean distance on average. The retrieval recall of our scheme is about 5\% lower than the scheme retrieval by Euclidean distance on average.
Fig.~\ref{fig:f1} shows the F1-Measure of \textsf{MIPP} scheme and the scheme retrieving with Euclidean distance. The result shows the F1-Measure of our scheme is about 7\% lower than the scheme retrieving with Euclidean distance. As described in previous sections, our scheme supports multiple image owners and can preserve image similarity information in the cloud.
Therefore, the loss of retrieval accuracy of our scheme is the trade off these two aspects.

\begin{equation}
P = \frac{TP}{TP+FP}
\label{eq:precision}
\end{equation}

\begin{equation}
R = \frac{TP}{TP+FN}
\label{eq:recall}
\end{equation}

\begin{equation}
F1 = \frac{2PR}{P+R}
\label{eq:F}
\end{equation}

\subsection{Evaluation of time consumptions}
\begin{figure*}[!t]
\centering
\begin{minipage}[c]{0.48\textwidth}
\centering
\includegraphics[height=170px]{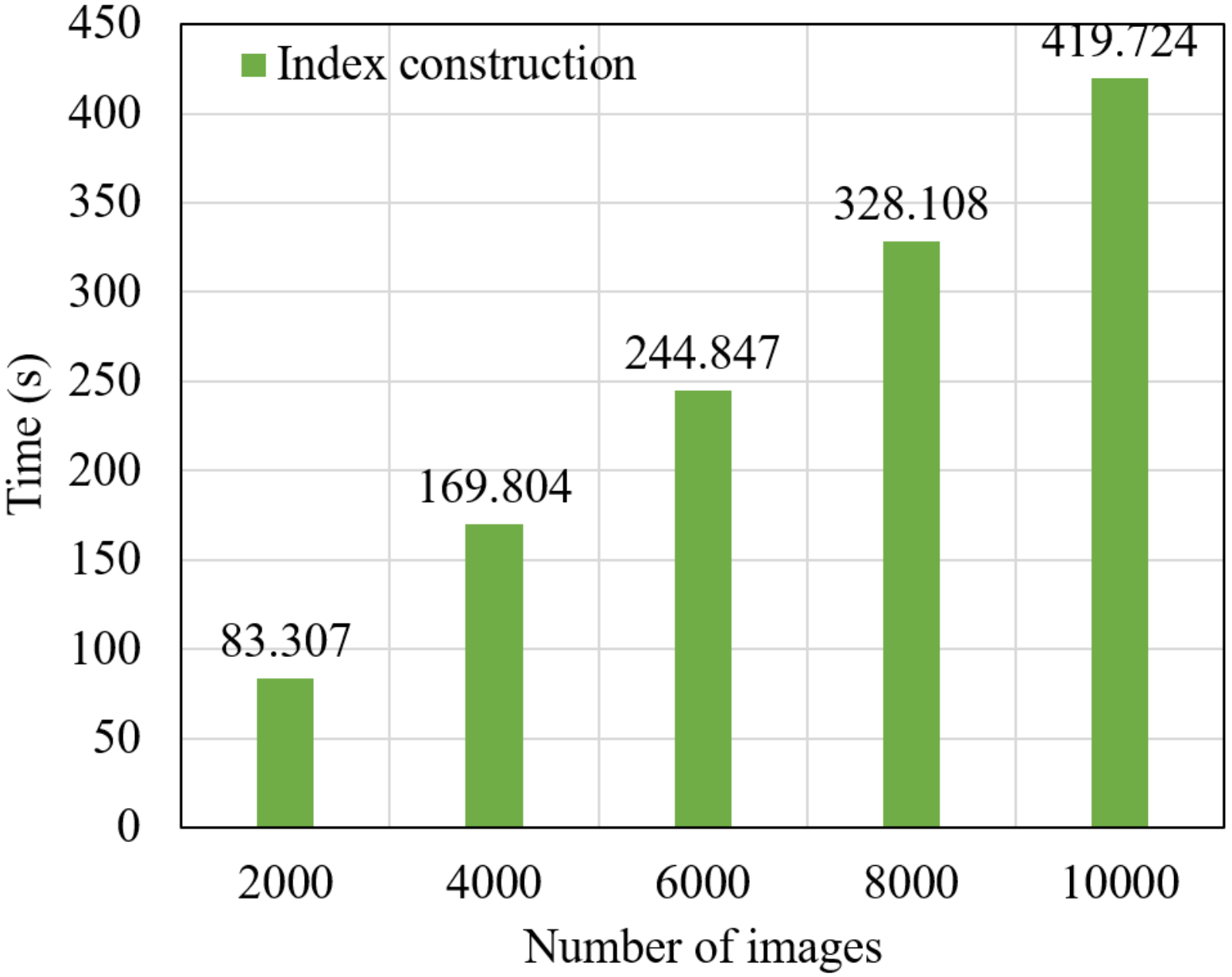}
\end{minipage}
\hspace{0.02\textwidth}
\begin{minipage}[c]{0.48\textwidth}
\centering
\includegraphics[height=170px]{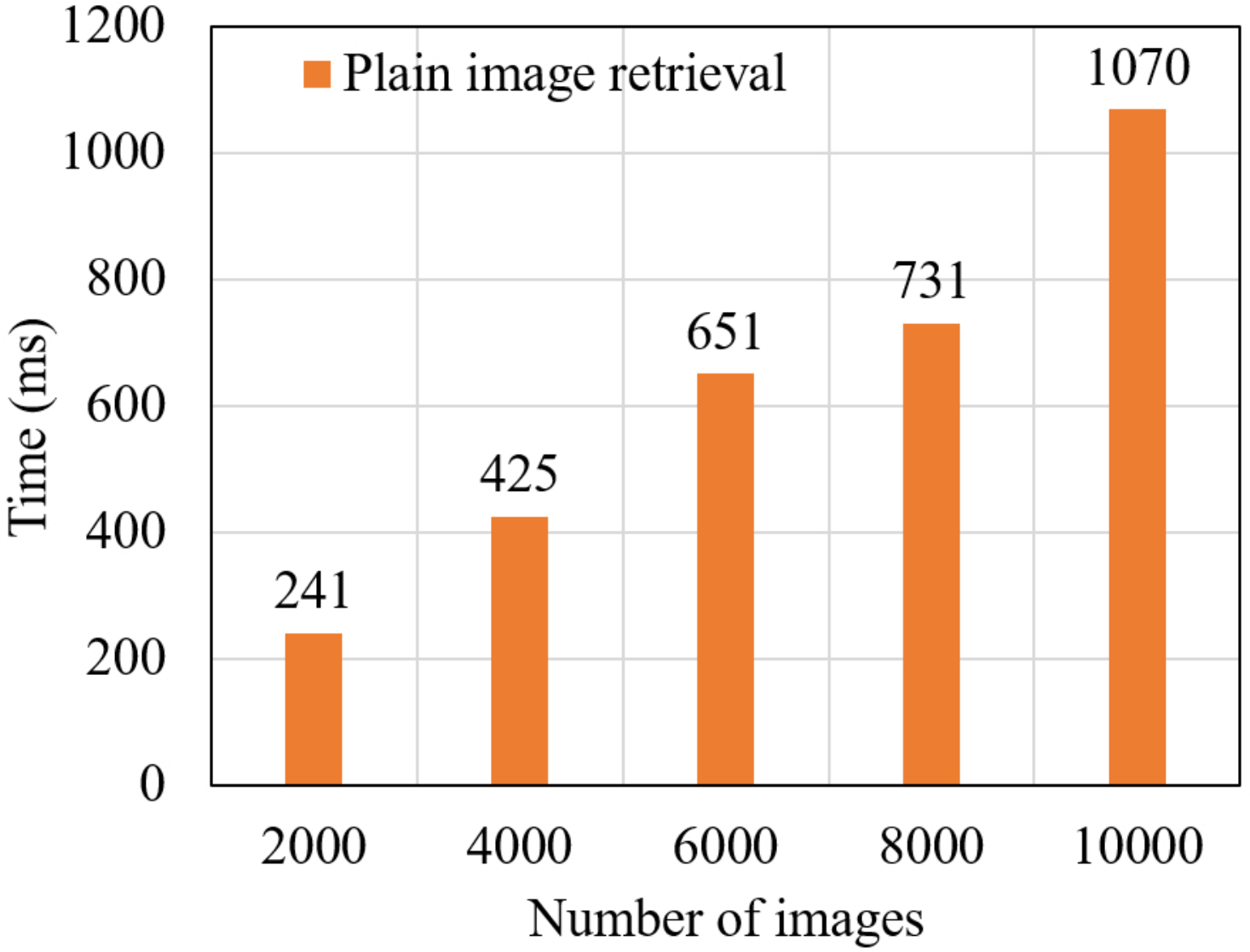}
\end{minipage}\\[3mm]
\begin{minipage}[t]{0.48\textwidth}
\centering
\caption{Index construction consumption}
\label{fig:indexcon}
\end{minipage}
\hspace{0.02\textwidth}
\begin{minipage}[t]{0.48\textwidth}
\centering
\caption{Plain image retrieval consumption}
\label{fig:plainre}
\end{minipage}
\end{figure*}

\begin{figure*}[!t]
\centering
\begin{minipage}[c]{0.48\textwidth}
\centering
\includegraphics[height=170px]{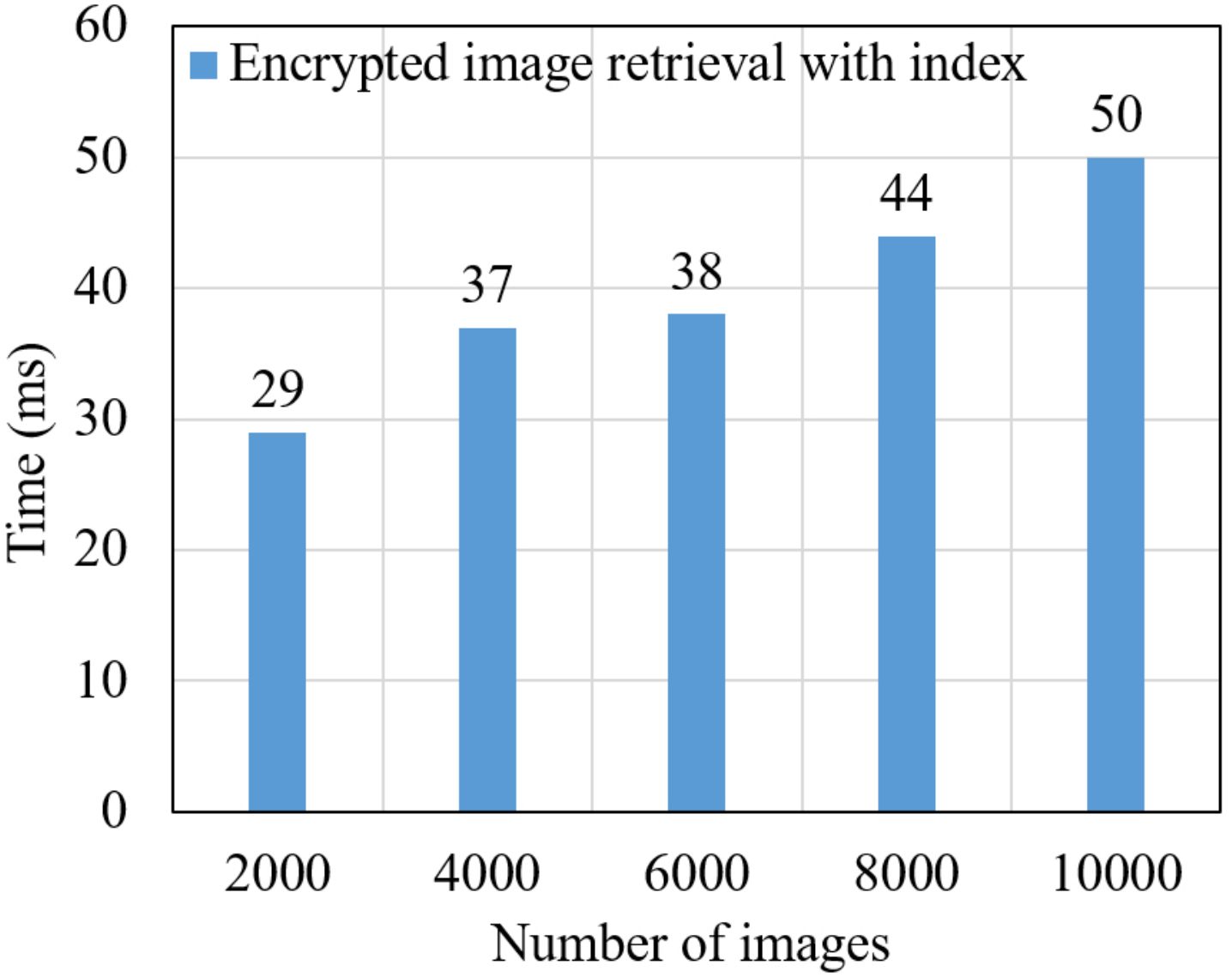}
\end{minipage}
\hspace{0.02\textwidth}
\begin{minipage}[c]{0.48\textwidth}
\centering
\includegraphics[height=170px]{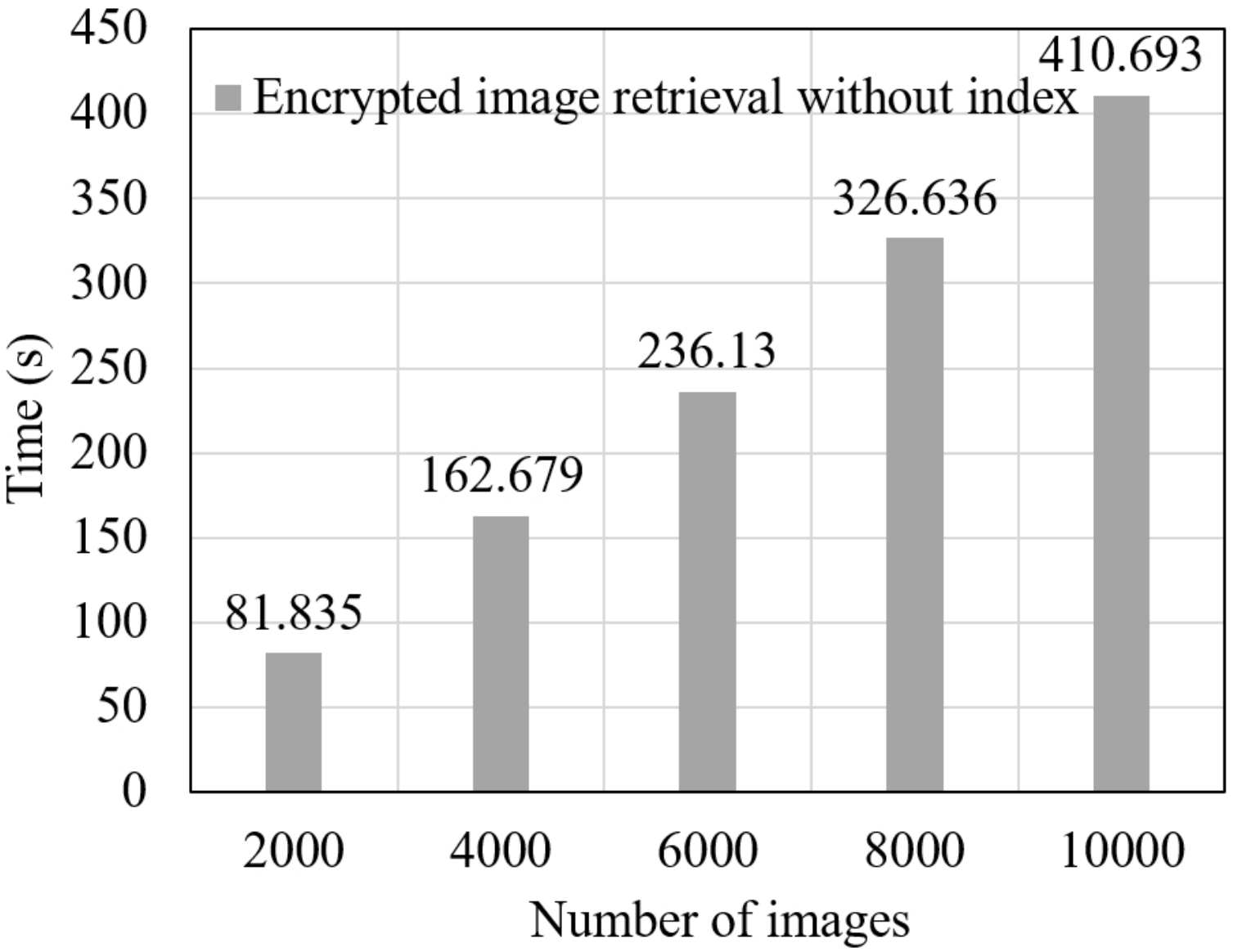}
\end{minipage}\\[3mm]
\begin{minipage}[t]{0.48\textwidth}
\centering
\caption{Consumption of encrypted image retrieval with index}
\label{fig:enreindex}
\end{minipage}
\hspace{0.02\textwidth}
\begin{minipage}[t]{0.48\textwidth}
\centering
\caption{Consumption of encrypted image retrieval without index}
\label{fig:enrenoindex}
\end{minipage}
\end{figure*}

The time consumptions of our scheme primarily consist of index construction time and secure image retrieval time, which is described as follows:

\begin{enumerate}[(1)]
\item Index construction time.

When a new image owner participates in our scheme, the image size will be increased. Fig.~\ref{fig:indexcon} shows the index construction time is increasing with the larger size of images. When the image size is 10,000, the index construction time is approximately 7 minutes, which is tolerable. After constructing the retrieval index, the efficiency of secure image retrieval in our scheme can be improved.

\item Secure image retrieval time.

The time consumptions of plain image retrieval, encrypted image retrieval with index, and encrypted image retrieval without index are shown in Fig.~\ref{fig:plainre}, Fig.~\ref{fig:enreindex}, and Fig.~\ref{fig:enrenoindex} respectively. The results show that the larger image size, the more time image retrieval consumes. The plain image retrieval time of 10,000 images consumes approximately 1 second. Encrypted image retrieval time without index of 10,000 images consumes approximately 6.8 minutes, while encrypted image retrieval time with index of 10,000 images consumes approximately 50ms. According to the above experimental data, we can calculate that index-based image retireval takes approximately 8,160 times faster than non-indexed image retrieval and index-based image retireval takes approximately 1,200 times faster than plain image retrieval when retrieving in a collection of 10,000 images. We can conclude that index-based encrypted image retrieval can greatly improve retrieval efficiency compared with plain image retrieval and non-indexed encrypted image retrieval.
The retrieval time-consuming result of index-based encrypted image retrieval shows that the retrieval efficiency of our scheme is appreciable.

\end{enumerate}

\subsection{Evaluation of storage consumption}
The storage consumption includes index storage and encrypted image features storage consumption, described as follows:

\begin{enumerate}[(1)]
\item Index storage consumption.

We extract image features to represent images for similarity calculation and construct retrieval indexes for improving retrieval efficiency. The storage consumption of index table is shown in Table 3. We can find that the retrieval index of 10,000 images costs approximately 267KB storage space. Therefore, the storage consumption of index table is very low.

\item Encrypted image features storage consumption.

In order to preserve image features privacy, we outsource encrypted image features to cloud. Table 3 shows that 10,000 encrypted image features consume approximately 3.2GB storage space. Because encrypted image features are stored in the cloud, which has high storage facilities, the storage consumption of encrypted image features in our scheme is tolerable.

\end{enumerate}

\begin{table}[!t]
\renewcommand{\arraystretch}{1.3}
\centering
\footnotesize
\tabcolsep 10pt %space between two columns. 用于调整列间距
\begin{tabular*}{8cm}{lll}
\multicolumn{3}{l}{\small{\textbf{Table 3}}}\\
\multicolumn{3}{l}{\small{Storage consumption of 10000 images.}}\\
\specialrule{0.05em}{3pt}{3pt}
   & Encrypted  & Retrieval  \\
   & Image Features & Index \\\hline
  Storage consumption (KB)& 3352278 & 267 \\
  \hline
\end{tabular*}
\label{tab:storage}
\end{table}

\section{Conclusion}\label{sec:conclusion}
In this paper, we presented a content-based multi-source encrypted image retrieval scheme in clouds with privacy protection.
We encrypted image features with the secure multi-party computation, which allowed image owners to encrypt image features by using their own keys. We also proposed a new method to measure the similarity of images that could avoid revealing image similarity information to the cloud at a certain extent.
Theoretical analysis and experimental results showed that our scheme enabled an accurate and efficient image retrieval over images gathered from multiple sources, while providing privacy guarantees. In the future work, we are to further improve the image retrieval efficiency.

\section*{Acknowledgements}
This work was supported in part by the National Science Foundation of China [Grant number 61602039] and the China National Key Research and Development Program [Grant number 2016YFB0800301].

\section*{References}

\bibliography{elsarticle}

\footnotesize

\vspace{20pt}

\begin{minipage}[b]{1.0\linewidth}
\textbf{Meng Shen} received the B.Eng degree from Shandong University, Jinan, China in 2009, and the Ph.D degree from Tsinghua University, Beijing, China in 2014, both in computer science. Currently he serves in Beijing Institute of Technology, Beijing, China, as an assistant professor. His research interests include privacy protection of cloud-based services, network virtualization and traffic engineering.
He received the Best Paper Runner-Up Award at IEEE IPCCC 2014.
He is a member of the IEEE. \\ \\
\end{minipage}

\begin{minipage}[b]{1.0\linewidth}
\textbf{Guohua Cheng} is a graduate student in the School of Computer Science, Beijing Institute of Technology.
Her research interests include image fusion, image retrieval, and privacy preserving algorithms. \\ \\
\end{minipage}

\begin{minipage}[b]{1.0\linewidth}
\textbf{Liehuang Zhu} is a professor in the Department of Computer Science at Beijing Institute of Technology.
He is selected into the Program for New Century Excellent Talents in University from Ministry of Education, P.R. China.
His research interests include Internet of Things, Cloud Computing Security, Internet and Mobile Security. \\ \\
\end{minipage}

\begin{minipage}[b]{1.0\linewidth}
\textbf{Xiaojiang Du} is a tenured professor in the Department of Computer and Information Sciences at Temple University, Philadelphia, USA.
His research interests are wireless communications, wireless networks, security, and systems.
He has authored over 230 journal and conference papers in these areas, as well as a book published by Springer.
Dr. Du has been awarded more than \$5 million US dollars research grants from the US National Science Foundation (NSF), Army Research Office, Air Force Research Lab, NASA, the State of Pennsylvania, and Amazon.
He serves on the editorial boards of three international journals.
Dr. Du is a Senior Member of IEEE and a Life Member of ACM.
\\ \\
\end{minipage}

\begin{minipage}[b]{1.0\linewidth}
\textbf{Jiankun Hu} is a Professor at the School of Engineering and IT, University of New South Wales (UNSW) Canberra (also named UNSW at the Australian Defence Force Academy (UNSW@ADFA), Canberra, Australia).
He is the invited expert of Australia Attorney-General’s Office assisting the draft of Australia National Identity Management Policy.
Prof. Hu has served at the Panel of Mathematics, Information and Computing Sciences (MIC), ARC ERA (The Excellence in Research for Australia) Evaluation Committee 2014.
His research interest is in the field of cyber security covering intrusion detection, sensor key management, and biometrics authentication.
He has many publications in top venues including IEEE Transactions on Pattern Analysis and Machine Intelligence, IEEE Transactions on Computers, IEEE Transactions on Parallel and Distributed Systems (TPDS), IEEE Transactions on Information Forensics \& Security (TIFS), Pattern Recognition, and IEEE Transactions on Industrial Informatics.
He is the associate editor of the IEEE Transactions on Information Forensics and Security.
\end{minipage}
\end{document}